\documentclass[pre,twocolumn,a4paper,showpacs]{revtex4}
\usepackage{natbib}
\usepackage{graphicx,pstricks}
\usepackage{epsfig}
\begin{document}

\title{A structural model of emotions of cognitive dissonances}
\author{Jos\'e F. Fontanari}
\affiliation{Instituto de F\'{\i}sica de S\~ao Carlos,
  Universidade de S\~ao Paulo,
  Caixa Postal 369, 13560-970 S\~ao Carlos SP, Brazil}
\author{Marie-Claude Bonniot-Cabanac and Michel Cabanac}
\affiliation{Department of Psychiatry \& Neurosciences, Faculty of Medicine, Laval University, Quebec, Canada}   
\author{Leonid I. Perlovsky}
\affiliation{Athinoula A. Martinos Center for Biomedical Imaging, Harvard University, Charlestown, MA  and Air Force Research 
Laboratory,   Wright-Patterson Air Force Base, OH}

\begin{abstract}

Cognitive dissonance is the stress that comes from holding two conflicting thoughts  simultaneously  in the mind, usually arising when people are asked to choose between two detrimental or two beneficial options. In view of the well-established role of emotions in decision making, here we investigate whether the conventional structural models used to represent the relationships among basic emotions, such as   the Circumplex model of affect, can describe the emotions of cognitive dissonance as well. We presented a questionnaire to 34 anonymous participants, where each question described a decision to be made among two conflicting motivations and asked the participants to rate analogically the   pleasantness   and the intensity of  the experienced emotion.  We found that the results were compatible with the predictions of the Circumplex model for basic emotions.

\end{abstract}
\maketitle

\section{Introduction} \label{sec:intro}

The notion of cognitive dissonance as the unpleasant motivational  state that  results from the inconsistency between people's behaviors and cognitions was put
forward by the Stanford psychologist Leon  Festinger about five decades ago \citep{Festinger_57}. As first noted by Festinger, to reduce this dissonance people seek to rationalize their
behaviors by overvaluing their choices and undervaluing the rejected alternatives. 

The recognition that cognitive dissonance plays a
key role  in people's behavior when choosing between alternatives led to the introduction  of the so-called free-choice paradigm \citep{Brehm_56}: since the
selected  alternative is unlikely to be perfect, and the rejected one is likely to have some desirable properties, making an irreversible choice 
between them   leads to the feeling
of discomfort associated to cognitive dissonance. Interestingly, the literature on the free-choice problem has focused exclusively on the \textit{post-decision} changes
in the assessment of  the  values of the alternatives, i.e., overvaluing our choices \citep{Brehm_56, Festinger_64,Gerard_83,Shultz_99,Chen_10}, a finding that is closely related to the basic
human bias of overestimating what we own, the so-called endowment effect \citep{Kahneman_91}.

In this contribution we begin the exploration of a different research vein, namely, the characterization of the emotions people feel at the very  moment they
are prompted to  make a decision or to choose between two qualitatively different alternatives -- these are the emotions of cognitive dissonances. 
Of course, the quantitative characterization of emotions -- whether associated to cognitive dissonances or not -- is itself a major research problem to which 
there is no  consensual solution at the moment \citep{Russell_99}. Here we follow Russell's suggestion   that whenever a measure of emotion
is needed one should use scales of pleasure-displeasure and arousal-sleepiness \citep{Russell_89}. Accordingly, we have presented a questionnaire containing  10 choice-questions  
to 34 participants and asked them to rate the intensity and the hedonicity (pleasantness) of the emotions elicited by those choices. In a second part of 
the experiment, we ask the participants to write a single emotion word that best describes the nature of the experienced emotion.

The main drawback of our experimental procedure is that by registering  the degrees of arousal and pleasantness  only, we discard \textit{a priori } other
dimensions that may also be important to characterize the emotions of cognitive dissonance. Nevertheless, even within this limited scenario we can
test whether those emotions can be described by the Circumplex model of affect,   in which emotions are arranged in a circular form with two bipolar dimensions interpreted 
as the degree of pleasure and the degree of arousal \citep{Russell_80,Cabanac_02}. In that sense, emotions mix together in a continuous manner like hues around the color circle \citep{Russell_89}.
A reorientation and, consequently, reinterpretation of the axes of the Circumplex model as positive affect and negative affect has been suggested to correct for the
 fact that there were few emotions in the neutral middle region of the pleasantness-unpleasantness axis \citep{WT_85}.  

We found that the measures of arousal $E$  and pleasantness $H$ obtained from the questionnaires are not independent quantities, contrary to the  prediction of 
the Circumplex model.  
Most remarkably, however, we found that the axes determined by the directions of the first and second principal components of the matrix of  data were in  fact associated
to  actual dimensions of pleasantness and arousal according to the emotion words  used by the participants. In addition,  the central region of the 
$ \left ( E, H \right )$   plane  where the Circumplex model predicts emotions should be absent is fittingly described by the word \textit{indecision} by
the majority of the participants. Hence in the context of the experiment reported  here we conclude that  our characterization of the emotions of
cognitive dissonance is consistent with the predictions of the Circumplex model.

The rest of this paper is organized as follows.  In Sect.\ \ref{sec:Method}  we describe the procedure used to apply the questionnaires with the
choice questions to the participants. The items of the questionnaires are presented in the Appendix. In  Sect.\ \ref{sec:res} we present a statistical
study of the answers to the choice questions, emphasizing the differences due to the gender of the participants. The correspondence between emotion words and
the regions in the $ \left ( E, H \right )$   plane is obtained using a clustering algorithm and the suitability of the Circumplex model to represent our data is
discussed in that section too.  Finally,  Sect.\ \ref{sec:Conc} summarizes our main conclusions. An abridged version 
of the present paper   was published in \citet{IJCNN_11}.

\section{Method} \label{sec:Method}

As is previous studies \citep{16,17,18,19,20,21,22,23,24}, mental experience was explored in interviews where participants answered printed questionnaires (see also \cite{25,26,27}). 
Thirty-four anonymous participants (who were referred to by numbers only), 17 men (age  $49 \pm 17$ yr.) and 17 women (age $50 \pm 17$ yr.) were presented 
two questionnaires each containing ten items.  Both questionnaires presented the same items, but the participant was asked to rate experienced pleasantness or hedonicity 
 ($H$) from one and intensity ($E$) from the other.  
All items described a decision to be made among two conflicting motivations and the participant was to rate analogically the magnitude of her/his experience. 
 
Questionnaire E explored emotion: a horizontal line was present below the item with a zero mark  on its left end. The participant was to pencil a small vertical mark at that line
 rating the intensity of the experienced feeling. The distance from the zero mark would indicate the magnitude of the experience, denoted by $E$.
  After rating the magnitude of the emotion, the participant wrote one 
 word describing the nature of the experienced emotion, e.g., curiosity, surprise, joy, indifference, anger, etc..  We were able to  obtain the emotion words from 33 of the 34 participants.
 
Questionnaire H explored hedonicity:  as before, a horizontal line was present below the item but with a zero on its middle, a minus (-) sign at the left end and a plus (+) sign at the right end.  
The participant was to pencil a small vertical mark on the right side of that line if the feeling was pleasant, or on the left side if unpleasant.  The distance from the middle (zero mark) of the 
line would indicate the magnitude of the experienced hedonic feeling, denoted by $H$.

Thus, the hedonic and magnitude feelings were measured quantitatively in millimeters, as well as recorded semantically. In order to minimize a possible influence of answering one questionnaire
 on the response to the other questionnaire, Questionnaires E and H were presented separately over time spans that varied from about one hour to half a day, depending on  the availability of the participant; 
 17 of them received Questionnaire E first, then Questionnaire H and the other 17 started with Questionnaire H. The first questionnaire was presented in the morning period with the care to keep the gender 
 of the participants balanced, and the second questionnaire was presented in the afternoon.

The ten items describing decisions to be made covered a broad range of motivations, from minor decisions in the daily life (e.g., how about movie or theater for tonight?) to clear but non-vital problems 
(e.g., would you go for a high-gain but risky investment or for a low-gain but secure one?) 
and finally to vital problems (e.g., would you go for radical surgery or for life-long therapy to treat a severe illness?). The ten items are presented in the appendix.

\section{Results}\label{sec:res}

Our analysis of the answers to the questionnaire items  is greatly facilitated by the fact that they can be represented in a two dimensional arousal-pleasantness
$(E,H)$  graph. So we begin our study by presenting a scatter plot showing the raw data (Sect.\ \ref{SP})  and then proceed to a more detailed account of the gender-dependent  distribution of 
answers for each choice question (Sect.\ \ref{QD}).  The assignment of emotion names to each item of the questionnaire  allows us to use those names to tag points in the  $(E,H)$ plane and then
define the distances between emotion words as the Euclidean distance between points in that plane. Given these distances we use a hierarchical clustering algorithm to partition the
emotion names into 8 categories (Sect.\ \ref{Emotion}). A summary of the main results is presented in Sect.\ \ref{Discussion}.

\subsection{Scatter plot}\label{SP}

Our first task is to turn the analogical  measures $E$ and $H$ into dimensionless quantities. Recalling that the degree of arousal $E$ takes on positive values only and the degree of pleasantness $H$ takes on positive as well as negative values,  we can rescale these measures by their maximum and minimum values so that  $E \in \left [ 0,1 \right ]$ and 
$H \in \left[ -0.5,0.5 \right ]$, without loss of generality.  In order to facilitate the visual inspection of the spread
of these quantities in a two dimensional graph,  we have equated the sizes of the  domains of  $E$ and $H$. 

\begin{figure}
\centerline{\epsfig{width=0.47\textwidth,file=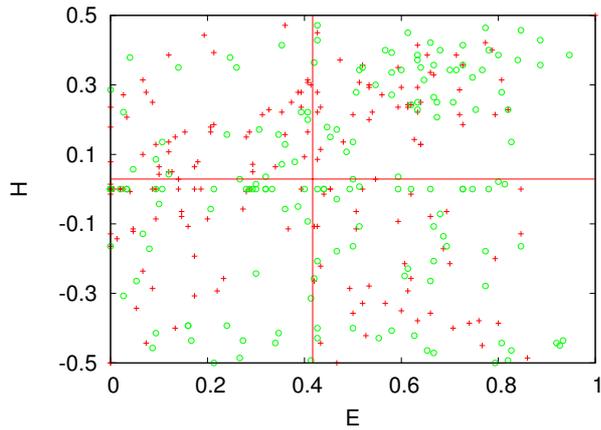}}
\par
\caption{Scatter plot of the degrees of arousal  and pleasantness. The crosses indicate the  arousal-pleasantness coordinates obtained from 
men's rates  and the open circles from  women's.
The horizontal straight line indicates the location of the mean value
$\bar{H} = 0.029$, whereas the vertical indicates the location of $\bar{E} = 0.417$.}
\label{fig:1}
\end{figure}

In Fig.\ \ref{fig:1} we 
show that two-dimensional graph (scatter plot)  where the symbols indicate the values of the properly rescaled degrees of arousal  $E $ and pleasantness $H$ for the 340 points associated to the 
10 choice questions of the 34 participants, as described in the previous section.  We have separated the answers -- for simplicity we will refer to a coordinate  
$\left ( E,H \right )$  as an answer
to a corresponding choice question -- according to the gender of the participants so that the crosses in  Fig.\ \ref{fig:1} represent men's answers and the circles,  women's.
In particular, the mean degrees of arousal and pleasantness  associated to men's answers  are $\bar{E}_m = 0.380$ (standard deviation  $\sigma_m^E = 0.257$) and 
$\bar{H}_m = 0.038$  (standard deviation $\sigma_m^H = 0.245$), respectively, whereas for women's we find  $\bar{E}_f = 0.454$ (standard deviation $\sigma_f^E = 0.270$) and
$\bar{H}_f = 0.021$ (standard deviation $\sigma_f^H = 0.275$). Regardless of gender, these statistical measures yield 
 $\bar{E} = 0.417$ (standard deviation $\sigma^E = 0.260$) and   $\bar{H} = 0.029$ (standard deviation $\sigma^H = 0.260$), which are also presented in the scatter plot of Fig.\ \ref{fig:1}. Interestingly, in the average, 
 women  exhibited a higher degree of arousal but a lower degree of pleasantness than  men. The histograms 
exhibiting the distribution of $E$  and $H$  values were presented in \citet{IJCNN_11}; here we just mention that about $18\%$ of the $H$ values are very  close to its mean value
so $\bar{H}$ is actually the most likely value of $H$ (visual inspection of the scatter plot confirms this claim), whereas only about $5\%$ of the values of $E$ are very  close to 
its mean $\bar{E}$.

%
\subsection{Characterization of the  choice questions answers}\label{QD}
\begin{figure}
\centerline{\epsfig{width=0.47\textwidth,file=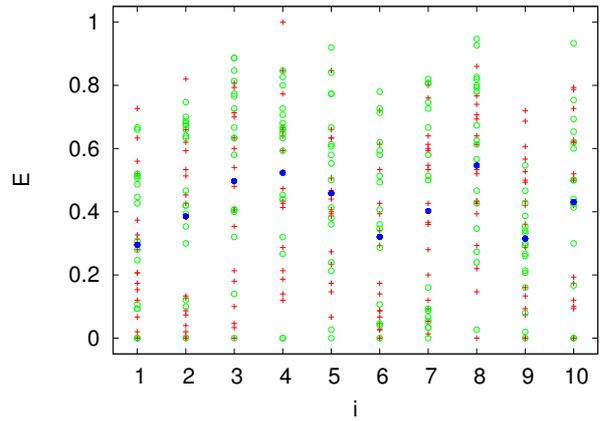}}
\par
\caption{The degree of  arousal $E$  for each of the $i=1,\ldots,10$ choice questions. The crosses are men's  answers and the open circles, women's. 
The filled circles indicate  the  mean degree of pleasantness  of each question regardless of gender.}
\label{fig:2}
\end{figure}

\begin{figure}
\centerline{\epsfig{width=0.47\textwidth,file=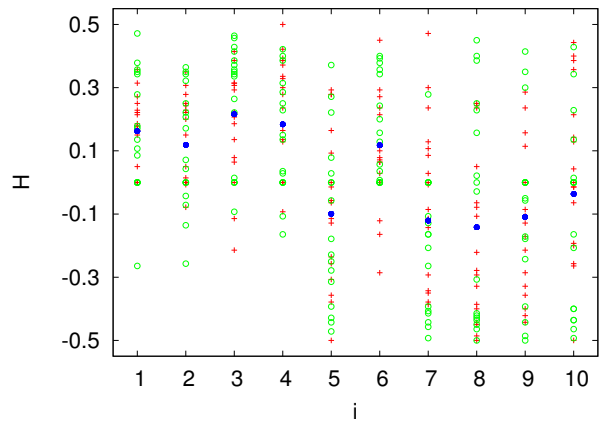}}
\par
\caption{The degree of pleasantness $H$   for each of the $i=1,\ldots,10$ choice questions. The crosses are men's answers and the open circles, women's.
The filled circles indicate  the  mean degree of arousal  of each question regardless of gender.}
\label{fig:3}
\end{figure}

In order to better acquaint the readers with the participants answers to the ten choice questions, we present in Figs.\ \ref{fig:2} and \ref{fig:3} the
degrees of arousal and pleasantness separated by gender for each question. In addition, Tables \ref{table1} and \ref{table2} exhibit the mean values of those
degrees; the standard deviations can be estimated by visual inspection of the figures. To appreciate the underlying structure of the participant answers we compare them
with a null model  in which  $E$ and $H$ are  chosen randomly and uniformly in the ranges $\left [ 0,1 \right ]$    and  $\left [ -0.5, 0.5 \right ]$. In this case the
null model mean degree of arousal associated to a given item is a  sum of 34 independent random variables uniformly distributed in $\left [ 0,1 \right ]$ and so it has mean $0.5$
and standard deviation  $1/\sqrt{34 \times 12} \approx 0.05$. The same is true for the null model mean degree of pleasantness except that its mean is zero.
Inspection of Tables \ref{table1} and \ref{table2} indicate that for some items  the range of variation of the participants' answers is far greater than that predicted by the null model.

%

\begin{table}
\caption{Mean degree of arousal for men ($\bar{E}_m$),  women $\bar{E}_f$ and gender-independent ($\bar{E}$)  for the
 ten choice questions. The last column  shows a sample of the null model.  } 
\centering 
\begin{tabular}{c c c c c }
\hline
~~ $i$ ~~ &  ~~ $\bar{E}_m$ ~~ & ~~ $\bar{E}_f$ ~~ & ~~ $\bar{E}$ ~~& ~ null model ~ \\ [0.5ex] 
\hline 
  1  & 0.275   &   0.314  &  0.294  &        0.514 \\
  2  & 0.300   &   0.470  &  0.385  &     0.636\\
  3  & 0.412   &   0.581  &  0.496  &     0.434 \\
  4  & 0.501   &   0.545  &  0.523  &       0.541  \\
  5  & 0.418   &   0.497  &  0.457  &       0.561 \\
  6  & 0.233   &   0.407  &  0.320  &       0.467  \\
  7  & 0.411   &   0.393  &  0.402  &    0.550 \\
  8  & 0.495   &   0.597  &  0.546  &    0.549\\
  9  & 0.356   &   0.273  &  0.314  &      0.522  \\
 10 & 0.399   &   0.461  &  0.430  &      0.456  \\  [1ex] 
\hline 
\end{tabular}
\label{table1} 
\end{table}

\begin{table}
\caption{Mean degree  of pleasantness for men ($\bar{H}_m$), women $\bar{H}_f$ and gender-independent ($\bar{H}$) for the
 ten choice questions. The last column  shows a sample of the null model. } 
\centering 
\begin{tabular}{c c c c c }
\hline
~~ $i$ ~~ &  ~~ $\bar{H}_m$ ~~ & ~~ $\bar{H}_f$ ~~ & ~~ $\bar{H}$ ~~& ~ null model ~ \\ [0.5ex] 
\hline 
  1  &     0.172   &     0.152  & 0.162  & -0.009 \\
  2  &   0.129     &     0.107  & 0.118  & -0.014 \\
  3  &     0.178   &     0.253  & 0.215  &  0.048\\
  4  &    0.228    &     0.139  & 0.183  &  0.069\\
  5  &   -0.077   &    -0.121  & -0.099  &  0.034\\
  6  &   0.076    &     0.158  & 0.117   &   0.072\\
  7  &   -0.043   &    -0.197  & -0.120  &  0.043\\
  8  &   -0.191   &    -0.091  & -0.141  &  0.019\\
  9  &   -0.132   &    -0.086  &  -0.109 &  0.039\\
 10 &    0.036   &    -0.108  &  -0.036  & -0.042\\  [1ex] 
\hline 
\end{tabular}
\label{table2} 
\end{table}

A more useful piece of information is the correlation between the degrees of arousal and pleasantness for each question. This
quantity can be calculated by introducing the item-dependent covariances
\begin{equation}\label{cov_1}
Cov^i \left ( E, H \right ) = \left \langle E H \right \rangle_ i - \left \langle E  \right \rangle_ i \left \langle  H \right \rangle_ i
\end{equation}
with $i=1,\ldots,10$ and is  $\left \langle X \right \rangle_ i = \sum_ {\alpha=1}^{34} X_i^\alpha /34 $ is an item-dependent expected value. 
Hence the measured correlation coefficient for item $i$    is
\begin{equation}\label{cor_1}
Cor^i \left ( E, H \right ) = \frac{Cov^i \left ( E, H \right ) }{\left [ Cov^i \left ( E, E \right ) Cov^i \left ( H, H \right )  \right ]^{1/2} }.
\end{equation}

Table \ref{table3} exhibits these
correlations together with a sample of correlations generated by the  null model described before.   Of course, the choice questions have no
influence on  the correlation values for the null model.

\begin{table}  
\caption{Correlation between the degrees of arousal and pleasantness for each choice question. The third column is a sample of
the null model.} 
\centering 
\begin{tabular}{c c c }
\hline
~~ $i$ ~~ &  ~~ $Cor^i \left ( E,H \right) $ ~~ & ~~ null model ~~ \\ [0.5ex] 
\hline 
  1  & 0.393     &    0.178  \\
  2  &  0.318    &    0.176  \\
  3  &  0.527   &   0.247  \\
  4  &  0.209     &    0.169  \\
  5  &  -0.157     &    -0.278 \\
  6  &  0.559     &    0.304   \\
  7  & -0.003  &  -0.  013  \\
  8  &   0.112 &    -0.138\\
  9  &  0.061 &     0.067 \\
 10 &   0.009 &    -0.001   \\  [1ex] 
\hline 
\end{tabular}
\label{table3}
\end{table}

To  facilitate the comparison with the null model we consider the mean correlation regardless of the choice question, which is
obtained by adding up the correlations in the second column of Table \ref{table3} and dividing the result by the number of items. We find that the
final result  $Cor \left ( E,H \right)  = 0.203$ is about  four standard deviations apart from the result predicted by the null model. More importantly,
the finding that this correlation is significantly different from zero shows that $E$ and $H$ are not independent quantities as assumed in the 
Circumplex model of affection. In fact, we recall that Questionnaires E and H were applied in different periods of the day exactly to  minimize the
influence of the answering one questionnaire on the response to the other questionnaire. Hence the correlation reported here is not an artifact of the
experimental setup.  We will return to this point in Sect.\ \ref{Discussion}.

\subsection{Emotion names}\label{Emotion}

As pointed out in Sect.\ \ref{sec:Method}, 33 participants described the emotions they felt at making a choice by a single emotion word. They used  a total of 77 different emotion words for the 330 choice questions. (See Table \ref{table6}  for a list of all emotion words used by the participants.)
In Table \ref{table4} we present the ten most frequently used emotion words together with their frequencies. In addition,
we note that 35  emotion words were used only once, and 13 were used twice. In \citet{IJCNN_11} we have lumped the 77 emotion words together  into 18 classes according to
our common sense intuition of the proximity between those words. 

\begin{table}  
\caption{The ten more frequently used emotion names together with their frequencies.} 
\centering 
\begin{tabular}{l c }
\hline
~~ emotion name~~ &  ~~ frequency ~~ \\ [0.5ex] 
\hline 
 indifference    &  48 \\
 joy   &     25  \\
 interest  &    22 \\
  pleasure  &     16  \\
  hope  &      13 \\
  expectation  &      13  \\
  desire &   11 \\
  anxiety  &      10\\
  fear  &      10 \\
 surprise &      9   \\  [1ex] 
\hline 
\end{tabular}
\label{table4}
\end{table}

The first issue we need to  address is  whether the participants used the 
77 emotion words  in a coherent way, i.e., whether different   participants used the same emotion word to describe their emotions for the same choice question. 
To quantify this expectation, we will calculate the probability that two randomly selected participants   describe their emotions by the same word  for a same choice question. 
The desired probability, which we denote by $P^i$, can be estimated by counting the number of pairs of participants (there are $528$ pairs in total) who choose the same emotion word  
for each choice question $i=1, \ldots,10$    and then dividing the result by 528. Table \ref{table5} shows these probabilities for the 10 items together  with a realization of a hypothetical situation in 
which the participants pick the 77 words
with probability proportional to their frequencies  (see Table \ref{table4} for the frequencies of the most used words). The abnormally high value of  $P^6$ is due to the fact
that 11 participants used the word \textit{indifference} to describe their feelings at choosing between a violin and a piano sonata. 

\begin{table}  
\caption{Probability $P^i$  that one selects two participants at random and they describe the emotions elicited by choice question $i$  by same emotion word.  
The third column is a sample of
the null model.} 
\centering 
\begin{tabular}{c c c }
\hline
~~ $i$ ~~ &  ~~ $P^i $ ~~ & ~~ null model ~~ \\ [0.5ex] 
\hline 
  1  &  0.0757  &    0.0445  \\
  2  &  0.0738  &  0.0568   \\
  3  &  0.0530  &   0.0321  \\
  4  &  0.0890    &  0.0246   \\
  5  &  0.0284    &  0.0587  \\
  6  &  0.1193  &     0.0321  \\
  7  &  0.0625 &  0.0662  \\
  8  &  0.0719  &  0.0416 \\
  9  &  0.0511 &  0.0454  \\
 10 &  0.0435  &  0.0340  \\  [1ex] 
\hline 
\end{tabular}
\label{table5}
\end{table}

A better appreciation of the difference between the data and the random null model is achieved by considering the probability that two participants selected at random use the 
same word to describe the emotion evoked by the same choice question, \textit{regardless} of the question. This probability is $P_d = \sum_{i=1}^{10} P^i/10 =   0.0668$.
For the purpose of comparison, the same procedure applied to the probabilities in the third column yields $P_ r =  0.0437$. The relevant question here is whether the value of  
$P_d$ could be replicated by some realization of the random null model. To investigate this possibility we have generated  $10^6$ realizations such as that shown in the third column of Table \ref{table5}  so as to calculate the mean  and the standard deviation of the probability distribution of  $P_r$. The results are  
$\left \langle P_r \right \rangle = 0.0425$ and $\sigma_r = 0.0025 $. Hence   $P_d$  is about 10 standard deviations away from $\left \langle P_r \right \rangle$, which means we can safely discard the possibility that the assignment of the emotion words to the choice questions were random.   

The association of emotion words to items of the questionnaires  
offers us an opportunity to investigate the underlying organization of the emotion name categories, a line of research that has been extremely
influential on the  quantitative characterization of emotions in the 1980s \citep{Russell_80,WT_85,Shaver_87,Russell_89}. See, however, \citet{Russell_99} for a reappraisal of the 
conclusions drawn from those studies. An important outcome of this research avenue was the finding that emotion words are highly interconnected and so saying that
someone is anxious is not independent of saying   that person is happy or sad \citep{Russell_89}.  A complementary approach to the structural models of emotion names categories, such as the Circumplex model, is the exploration of the hierarchical structure of those categories \citep{Shaver_87}.

The central element  in those studies of emotion names categories is a distance matrix produced by asking individuals to rate the similarity between a given set of distinct emotion words using a fixed discrete scale. See \citet{Sergey_11} for an alternative approach where the distance is derived from the contexts in which the emotion names are used in web retrieved texts. Here we take advantage of our experimental setup described in Sect.\ \ref{sec:Method} to obtain an indirect measure of the distances between the emotion names used to describe the participants' feelings when answering the questionnaires items. More specifically, for each emotion word we associate a unique coordinate in the two dimensional space spanned by the
arousal and pleasantness dimensions. In the (typical) case where there are  $K$ points
$\left ( E^k,H^k \right ), k=1, \ldots, K$  associated to  the  same emotion word,  we associate that word to the mean  coordinate $\left ( \sum_k^K E^k/K ,\sum_k^K H^k/K \right )$.
Hence the distance between any two emotion words becomes simply the Euclidean distance between points in a plane.

The variance or spread of a set of points (i.e., the sum of the squared distances from the center) is the key element of many clustering algorithms \citep{Murtagh_97}.  
In Ward's minimum variance method \citep{Ward_63} we agglomerate two distinct clusters into a single cluster such that the within-class variance of the partition thereby obtained is minimal. Hence the method proceeds from an initial partition where all objects (77 emotion names, in our case) are isolated clusters and then begin merging the clusters so as to minimize the variance criterion. Table \ref{table6} shows the resulting partition of the emotion names into 8 clusters. We note that although these classifications are overall reasonable there 
are a few dissonant groupings  such as the lumping together of the words \textit{interest} and \textit{boredom} into category VI. However, as we will argue below the
appearance of antagonistic words in this particular category is somewhat expected since it describes indecision.  As each emotion word in Table \ref{table6} corresponds to
a point in the $(E,H)$ plane, it is natural to think of the location of the abstract categories in that plane as the position of the center of mass of the component words,
which are  presented in Table \ref{table7}.

\begin{table}  
\caption{The partition of the 77 emotion names into 8 clusters according to Ward’s minimum variance hierarchical clustering algorithm. } 
\centering 
\begin{tabular}{c l }
\hline
~~ category~~ &  ~~~~~~~~~~~~~~~~~ emotion names  ~~ \\ [0.5ex] 
\hline 
  I  &   joy, pleasure, delight, satisfaction, enthusiasm,   \\
& excitement, elation, greed, waiting, relaxation,  \\
&     relief, thinking, frustration,  despair,  challenge, \\
&   commitment,  curiosity. \\
  II  &  uneasiness,  puzzling, irritation, anxiety, distress,\\
& sadness,  indignation, hesitation, disgust, solidarity.\\
  III &  fun, indifference,  anticipation, rejection, comfort, \\
& patience, nervousness, difficulty, disdain.      \\      
  IV  &  displeasure, purpose,  wrath,  fatalism, weariness,\\
&  stress, unbelief, discouragement.   \\           
  V  &  well-being,  luck,  desire,  impatience, surprise, hope, \\
&  nostalgia,   courage,  expectation.       \\      
  VI  &  discomfort, embarrassment, guilt, anguish, interest, \\
&  incertitude, motivation, serenity, safety,  concern,\\
&   fear, doh!, indecision, swindle, anger,  contempt,\\
&    boredom.    \\
  VII  &  disarray,   furor,   exasperation.   \\
  VIII  &  uncertainty,  disappointment, perplexity,  repulsion.  \\  [1ex] 
\hline 
\end{tabular}
\label{table6}
\end{table}

\begin{table}  
\caption{The center of mass of each emotion name category in the arousal-pleasantness plane.} 
\centering 
\begin{tabular}{c c c }
\hline
~~ category~~~~ &  ~~~~~~~~~~ $E$ ~~~~~~~~~~~ & ~~~~~ $H$~~ \\ [0.5ex] 
\hline 
  I  &    $0.57 \pm  0.08$   &     $0.27 \pm 0.09$   \\
  II  &   $0.67 \pm  0.09$   &     $-0.19 \pm 0.05$\\
  III &   $0.10 \pm  0.05$   &     $0.01\pm 0.1$ \\
  IV &   $0.21 \pm  0.06$    &     $-0.39 \pm 0.08$ \\
  V  &   $0.39 \pm  0.05$    &     $0.16 \pm 0.03$\\
  VI  &  $0.42 \pm  0.09$    &      $-0.04 \pm 0.07$  \\
  VII  &  $0.78 \pm  0.05$   &     $-0.44 \pm 0.03$\\
  VIII &   $0.50 \pm  0.06$  &   $-0.35 \pm 0.05$\\ [1ex] 
\hline 
\end{tabular}
\label{table7}
\end{table}

\subsection{Discussion}\label{Discussion}

\begin{figure}
\centerline{\epsfig{width=0.47\textwidth,file=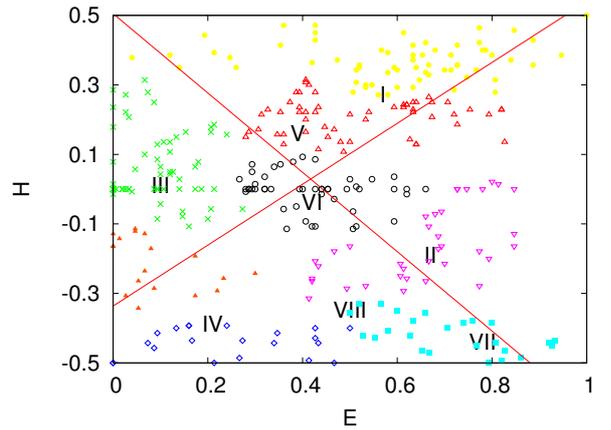}}
\par
\caption{Partition of the 340 arousal-pleasantness coordinate points into 8 clusters using Ward’s minimum variance hierarchical clustering algorithm.  Points belonging to different
clusters are represented by a different symbols. The orthogonal straight lines are the first and second principal components. The Roman numerals 
indicate the location of the emotion name categories  described in  Table \ref{table6}. }
\label{fig:5}
\end{figure}

To conclude our analysis we partition the 340  points in the scatter plot of Fig.\ \ref{fig:1}  into 8 clusters using Ward's minimum variance hierarchical clustering algorithm.
Note that, except for the emotion words used only once and so represented by a single point in the scatter graph,  this partition is different from the clustering of the 77 words
into the  8 categories summarized in Table \ref{table6}. The resulting partition together with the location of the emotion name categories given in Table \ref{table7}  are
exhibited in Fig.\ \ref{fig:5}. That figure together with Table \ref{table6} allow us to offer an interpretation for the 8  emotion name categories, namely,
I (pleasure), II (uneasiness), III (indifference), IV (displeasure), V (desire), VI (indecision), VII (furor),  and  VIII (disappointment). Of course, although there is a
considerable degree of arbitrariness in the naming of these categories we have chosen names that are representative of the \textit{majority} of the member words of
a category.

A few words are in order about Fig.\ \ref{fig:5} which summarizes the main  results of our analysis of the participants' answers to the items of the
questionnaires presented in the Appendix. Category VI is located at the center of the $(E, H)$ plane and so correspond to answers which are inconsiderable
from both the arousal and the pleasantness dimensions. In addition, the list of emotion words used to describe those answers comprehends pairs of antagonistic words such as
\textit{discomfort} and \textit{serenity},  and \textit{interest} and \textit{boredom}. Overall there is a well-balanced mixture of positive and negative emotion words which cancel out and
in the average one get neutral words such as \textit{incertitude} and \textit{indecision} which we think provide a very good description for category VI: it is not
associated to any particular emotion name. This is a most interesting situation because one of the predictions of the Circumplex model of affect is exactly an empty region
in the center of the $(E, H)$ plane \citep{Russell_80,Russell_89}. Another interesting point, is that classes I (pleasure) and IV (displeasure) are diametrically opposed, though not along
the $H$ axis as one would expect. 

As pointed out in Sect.\ \ref{QD}, our finding that the measures $E$ and $H$ are correlated, in spite of the experimental effort to minimize their influences on each
other, prompts us to look for a set of uncorrelated variables to describe the experimental points of the scatter plot shown in Fig.\ \ref{fig:1}. This is easily achieved
using principal component analysis  (PCA) and the results are exhibited by the orthogonal straight lines in Fig.\ \ref{fig:5}. The first principal component has 
the slope  $0.876$ and the second has the slope $-1.141$, so they are really orthogonal; they look distorted because the figure is not a square. Most interestingly,
the principal component corresponds  to the effective  dimension of pleasantness since categories I and IV are roughly located at its opposite extremes. The interpretation of the
second principal component is more difficult. Categories VII (furor) and II (uneasiness) fall very close to that axis which seems to represent a decrease in arousal (furor is more intense 
than uneasiness) but the lack of points in the other extreme of this axis prevents a more assertive interpretation. 

In summary, given the PCA reorientation of the axis and the interpretation of category VI as a `non-emotion' class  we found that our characterization of
cognitive dissonance emotions  is consistent with the Circumplex model of affect.

\section{Conclusion}\label{sec:Conc}

Decision-making in situations of conflicting motivations (cognitive dissonance) is a source of emotion,  usually described as a feeling of discomfort  that results from 
holding two conflicting thoughts  simultaneously  in the mind.
These decisions appear to be made in the hedonic dimension of consciousness; the hedonic experience taking place as an actual or an expected reward.  In this paper we made a step toward exploring a new type of emotions, aesthetic emotions related to knowledge or more specifically, emotions of cognitive dissonance related to contradictions between two pieces of knowledge. These emotions could  \textit{in principle} be different  from basic emotions. Whereas specific words exist to name basic emotions, there are no specific words for most emotions of cognitive dissonance. This fact might be a reason that these emotions have not been systematically studied in the psychological literature. 

Although the expression `cognitive dissonance' has been used for a long time \citep{Festinger_57, Brehm_56,Festinger_64}, emotions of cognitive dissonance have not been recognized as a special type of emotions different in principle from basic emotions. By presenting to participants questions as alternative mental choices, our paper 
presents the first steps to address  the intricate issue of distinguishing experimentally between aesthetic and basic emotions.

On the one hand,  it can be argued that there  is a fundamental theoretical difference between basic and aesthetic emotions. Following \citet{Grossberg_87}, basic emotions
can be considered  as feelings and mental states related to neural signals, which indicate to various brain regions satisfaction or dissatisfaction of fundamental organism needs. Mechanisms measuring these needs we call instincts. Hence basic emotions are mostly related to bodily needs, whereas aesthetic emotions are related to the need for knowledge.  In addition, 
\citet{7}  argues that emotions of cognitive dissonance could be in some way similar to musical emotions.

On the other hand, the experimental study reported here failed to uncover any distinction between basic and cognitive dissonance emotions; rather we found that
the latter can be described remarkably well by the Circumplex model, which is a structural model proposed to describe basic emotions 
\citep{Russell_80,Russell_89}.
It might well be that our experimental setup centered on the record of the degrees of arousal $E$ and pleasantness $H$ elicited by the choice questions  is not  sensitive enough for the  fine distinctions required to differentiate details of aesthetic emotions. The measurement of other emotion dimensions in addition to $E$ and $H$ may  be necessary for 
achieving that fine distinction, if indeed it exists.

To conclude, we note that understanding the underlying psychological structure of emotions is germane for the  development of robotic systems capable of exhibiting as well as
recognizing emotion-like responses \citep{Taylor_05,Lola_05,Levine_07,Khashman_10}. In fact, according to our results and, more generally, in conformity with the predictions of the Circumplex model of affect \citep{Russell_80,Russell_89}, the
combination of two
quantities -- the degree of arousal $E$ and the degree of pleasantness $H$ -- can explain a large part of the spectrum of human emotional experience. Hence the design of artificial neural networks 
with sensors and estimators for these two quantities may be an efficient manner to mimic human-like emotion responses in machines.  The neural network models for decision making
 based on positive or negative affect directed at objects or potential actions \citep{Gutowsky_87,Leven_96} can be viewed as examples of work in this research direction.

\section*{Acknowledgments}

The research  at  S\~ao Carlos was  supported by  The Southern Office of Aerospace Research and Development (SOARD), grant FA9550-10-1-0006,  and 
Conselho Nacional de Desenvolvimento Cient\'{\i}fico e Tecnol\'ogico (CNPq).

\section*{Appendix}

The 10 items of Questionnaire E aiming at measuring the degree of arousal of the evoked  emotion  are presented  below. We note that the  questions were
formulated in French (the participants were native French speakers), so the following items are a nonliteral translation of the original items.

\begin{enumerate}

\item[1] Focus on what you feel when you are asked to make the following choice:  Do you prefer red  or white wine to accompany
duck with orange? 
Do you feel an emotion at the idea of this choice? Indicate its intensity on the line below.

0 \line(1,0) {180}  Max

\item[2] Focus on what you feel when you are asked to make  the following choice: Do you prefer  cinema or theater? 
Do you feel an emotion at the idea of this choice? Indicate its intensity on the line below.

0 \line(1,0) {180}  Max

\item[3] Focus on what you feel when you are asked to make the following choice: Do you prefer the sea or the mountain for the holiday season? 
 Do you feel an emotion at the idea of this choice? Indicate its intensity on the line below.

0 \line(1,0) {180}  Max

\item[4] Focus on what you feel when you are asked to make the following choice: Do you prefer  to receive a large amount of money in a single parcel or the same amount in small parcels? 
 Do you feel an emotion at the idea of this choice? Indicate its intensity on the line below.

0 \line(1,0) {180}  Max

\item[5] Focus on what you feel when you are asked to make the following choice:
Do you prefer a secure but  relatively poorly paid job or a very well-paid job but at risk of loss of employment?
 Do you feel an emotion at the idea of this choice? Indicate its intensity on the line below.
 
0 \line(1,0) {180}  Max

\item[6] Focus on what you feel when you are asked to make the following choice:
Do you prefer to hear  a violin    or a  piano sonata?
 Do you feel an emotion at the idea of this choice? Indicate its intensity on the line below.
 
0 \line(1,0) {180}  Max

\item[7] Focus on what you feel when you are asked to make the following choice:
If your employer requires you to learn a Scandinavian language, which one would you  prefer,  Norwegian or Swedish? 
 Do you feel an emotion at the idea of this choice? Indicate its intensity on the line below.
 
0 \line(1,0) {180}  Max

\item[8] Focus on what you feel when you are asked to make the following choice:
In order to treat a serious illness would you opt for a quick surgery or for a life-long therapy?
 Do you feel an emotion at the idea of this choice? Indicate its intensity on the line below.
 
0 \line(1,0) {180}  Max

\item[9] Focus on what you feel when you are asked to make the following choice:
Do you prefer a  comprehensive but very expensive insurance  or a cheaper one  but with many  gaps?
 Do you feel an emotion at the idea of this choice? Indicate its intensity on the line below.
 
0 \line(1,0) {180}  Max

\item[10] Focus on what you feel when you are asked to make the following choice:
Would you vote for a  right wing  party, which guarantees citizen's security, or for a left wing party, which promotes an egalitarian society?
 Do you feel an emotion at the idea of this choice? Indicate its intensity on the line below.
 
0 \line(1,0) {180}  Max

\end{enumerate}

In order to measure the degree of pleasantness felt by the participants in making those choices, they were presented the same ten-questions questionnaire  and 
after reading each item they were asked to rate their pleasure by penciling a small vertical mark on  the straight line

-  \line(1,0) {102}  ~ 0 \line(1,0) {102}  +.

Here the  -  sign indicates a most unpleasant choice and the + sign a most pleasant one, and the distance from zero is 
the analog magnitude rating of hedonicity.  This set of questions comprises Questionnaire H as described in Sect.  \ref{sec:Method}.

\end{document}